\documentclass[a4paper,11pt]{article}
\pdfoutput=1 

\usepackage{jcappub} 

\usepackage[T1]{fontenc}

\def\lsim{\mathrel{\rlap{\lower4pt\hbox{\hskip1pt$\sim$}} \raise1pt\hbox{$<$}}}
\def\gsim{\mathrel{\rlap{\lower4pt\hbox{\hskip1pt$\sim$}} \raise1pt\hbox{$>$}}}

\newcommand\ahigh{\mathrel{\overset{\makebox[0pt]{\mbox{\normalfont\tiny\sffamily High}}}{\approx}}}

\usepackage{hyperref}
\usepackage{graphicx,xcolor}
\usepackage[export]{adjustbox}
\usepackage[normalem]{ulem}
\usepackage{cancel}

\definecolor{grassgreen}{cmyk}{0.77,0,1,0.05}

\usepackage{slashed}
\allowdisplaybreaks
\usepackage{tensor}

\newcommand{\half}{\frac{1}{2}}
\newcommand{\be}{\begin{equation}}
\newcommand{\ee}{\end{equation}}
\newcommand{\bea}{\begin{eqnarray}}
\newcommand{\eea}{\end{eqnarray}}
\newcommand{\nn}{\nonumber}
\newcommand{\parent}[1]{\left(#1\right)}
\newcommand{\cL}{\mathcal{L}}
\newcommand{\Mpbar}{\overline{M}_{\text{Pl}}}
\newcommand{\Mp}{M_{\text{Pl}}}

\newcommand{\eq}[1]{Eq.~\eqref{#1}}
\newcommand{\Figref}[1]{Fig.~\ref{#1}}
\newcommand{\Tabref}[1]{Tab.~\ref{#1}}

\newcommand{\rsw}{\rm sw}
\newcommand{\turb}{\rm turb}
\newcommand{\coll}{\rm col}

\def\yj{\textcolor{violet}}

\title{Gravitational waves from a first-order phase transition of the inflaton}

\author[a,b]{Jörn Kersten,}

\emailAdd{jkersten@yonsei.ac.kr}

\author[a,c]{Seong Chan Park,}

\emailAdd{sc.park@yonsei.ac.kr}

\author[a]{Yeji Park,}

\emailAdd{yeji.park@yonsei.ac.kr}

\author[a]{Juhoon Son,}

\emailAdd{psiomega@yonsei.ac.kr}

\author[d,e]{Liliana Velasco-Sevilla}

\emailAdd{lilianak@sogang.ac.kr}

\affiliation[a]{Department of Physics and IPAP, Yonsei University, 
\\Seoul 03722, South Korea}

\affiliation[b]{Department of Physics and Technology, University of Bergen, \\Postboks 7803, 5020 Bergen, Norway}

\affiliation[c]{School of Physics, Korea Institute for Advanced Study, \\85 Hoegi-ro, Seoul 02455, South Korea}

\affiliation[d]{Center for Quantum Spacetime, Sogang University, 35 Baekbeom-ro, 
\\Seoul 121-742, South Korea}

\affiliation[e]{Department of Physics, Sogang University, 
\\35 Baekbeom-ro, Seoul 121-742, South Korea}

\abstract{We explore the production of gravitational waves (GW) resulting from a first-order phase transition (FOPT) in a non-minimally coupled `Dark Higgs Inflation' model. Utilizing a dark sector scalar field as the inflaton, we demonstrate how inflationary dynamics naturally set the stage for observable FOPT. These transitions, influenced by thermal and quantum effects, generate GW spectra potentially detectable by observatories such as LISA, DECIGO, the Cosmic Explorer and the Einstein Telescope. Our study highlights the inflaton's dual role in cosmic inflation and early Universe phase transitions, presenting a unified framework to probe physics beyond the Standard Model through gravitational wave astronomy.  }

\begin{document}
\begin{flushright}
CQUeST-2024-0752
\end{flushright}
\maketitle
\flushbottom

\section{Introduction}
\label{sec:intro}

The earliest epochs of the Universe are hidden from direct observation
by the opaqueness of the plasma before the formation of the Cosmic
Microwave Background radiation (CMB), blocking photon signals from
earlier times. Although we aim to detect more penetrating neutrino
signals, even neutrinos were in thermal equilibrium and scattering
frequently at temperatures above a MeV, preventing us from accessing
earlier information. Energy-wise, we are confined by the TeV cap
established by the LHC at present, and thus we cannot indirectly access
information about the early Universe above the TeV scale. It is,
therefore, crucial to explore alternative approaches to detect early
Universe signals instead of depending on the slowly advancing energy
capacities of colliders alone. Gravitational waves (GW) could be
a promising option.

Gravitational wave signals might arise from a variety of early Universe
sources \cite{Roshan:2024qnv}.  They could originate from the
inflationary era as well as from a violent first-order phase transition
(FOPT), for instance. The characteristics of such waves will vary
depending on their sources, providing opportunities to explore physics
in the early Universe beyond the TeV scale. This paper seeks to present
a unified framework for cosmic inflation alongside a FOPT, both stemming
from the same inflaton field. The observation of a GW signal from this
FOPT would be a vital tool for probing the early Universe and offer
crucial insights into inflation.

Among the various models for cosmic inflation, Higgs Inflation \cite{Bezrukov:2007ep} stands out as an appealing choice due to its simplicity and compatibility with observational data, distinguishing it from other well-known models such as chaotic inflation, power-law inflation, and natural inflation. The model requires only one modification to the Standard Model (SM): a non-minimal coupling of the Higgs doublet to the Ricci scalar, expressed as $\xi H^\dagger H R$, where $\xi$ is a dimensionless parameter~\cite{Futamase:1997fq, Park:2008hz}. From the perspective of effective field theory, this non-minimal coupling is natural, as quantum effects will generate it even if $\xi(\mu_c)=0$ is imposed by hand at a particular scale $\mu_c$. The normalization of the CMB
fixes the ratio between the non-minimal coupling parameter and the self-coupling constant at the scale of inflation:
\begin{align}
    \left.\frac{\lambda}{\xi^2}\right|_{\mu_{inf}} \sim 10^{-9} \,.
\end{align}
The hierarchy in the ratio is addressed in Critical Higgs Inflation~\cite{Hamada:2014iga,Hamada:2014wna}, where the self-coupling ($\lambda$) of the Higgs becomes small due to the significant influence of the large top quark Yukawa coupling on its renormalization group evolution. 
 
To evaluate the potential significance of the non-minimal coupling at low energy scales, we analyzed higher-dimensional operators, including D=6 and D=8, that emerge at these scales. However, their contributions are insufficient to induce a FOPT, consistent with the findings of \cite{Zhang:1992fs,Grojean:2004xa,Bodeker:2004ws,Delaunay:2007wb,Grinstein:2008qi,Damgaard:2015con,deVries:2017ncy,Chala:2018ari,Ellis:2019oqb} for EW scale GW production. 
On the other hand, the $R^2$ term can play a role in the ultraviolet (UV) completion of Higgs inflation models~\cite{Ema:2017rqn, Gorbunov:2018llf, Salvio:2015kka}. In these scenarios, the production of gravitational waves and primordial black holes can become significant due to the presence of a new scalar degree of freedom, known as the scalaron, introduced by the $R^2$ term~\cite{Cheong:2019vzl, Cheong:2022gfc}. In this paper, however, we focus on GW production through a FOPT in a single-field framework, without introducing the scalaron. 

In particular, we will consider FOPT occurring well below the inflationary energy scale down to the EW scale. Indeed, the interest in GW produced at the EW scale has exponentially increased due to the many planned and proposed experiments such as the Laser Interferometer Space Antenna (LISA) \cite{Colpi:2024xhw}, Einstein
Telescope \cite{Maggiore:2019uih}, Cosmic Explorer \cite{Evans:2021gyd},
DECIGO \cite{Kawamura:2020pcg} and BBO \cite{BBOref}
 that will be operating in the frequency ranges
from about $10^{-5}$\,Hz up to a few Hz, which is the characteristic frequency window associated with the EW scale GW production. 
To achieve a relatively low-scale FOPT, we extend beyond the SM Higgs inflation framework and consider `dark Higgs inflation.' This involves dark sector particles, including a dark sector scalar (the dark Higgs acting as the inflaton), dark sector fermions, and dark gauge bosons associated with a dark gauge symmetry. Numerous studies have shown that the breaking of dark gauge symmetry can induce an FOPT and generate observable GW~\cite{Azatov:2019png, Bhoonah:2020oov, Madge:2023dxc, Addazi:2023jvg, Arcadi:2023lwc, Chikkaballi:2023cce}.
Additionally, various models that incorporate dark Higgs fields and dark $U(1)$ symmetry breaking have been proposed~\cite{Hashino:2018zsi, Schwaller:2015tja, Jaeckel:2016jlh}, exploring energy scales ranging from sub-EW to TeV~\cite{Borah:2021ocu}. Some studies have specifically examined the dependence of GW frequencies on the symmetry-breaking scale~\cite{Dent:2022bcd, Bhoonah:2020oov}. While connections between inflation and phase transitions have been investigated~\cite{Ashoorioon:2015hya, Masina:2024ybn, Tenkanen:2016idg}, these works often rely on hybrid models or focus primarily on EW-scale FOPT.

Our work is structured as follows: Sec.~\ref{sec:u1} introduces the dark $U(1)_X$ model and its scalar potential, including thermal and one-loop corrections relevant to the phase transition. Sec.~\ref{sec:foptresults} presents benchmark points for a FOPT and demonstrates compatibility with CMB constraints on inflationary parameters. We conclude in Sec.~\ref{sec:conclusions}. Various technical details are discussed in appendices.

\section{Dark $\boldsymbol{U(1)}$ model \label{sec:u1}}

\subsection{Lagrangian} \label{sec:Lag}
In the minimal setup, the dark sector contains a complex scalar $\phi$
(``Dark Higgs'')
charged under a gauge group $U(1)_X$ and the corresponding gauge boson
$X^\mu$.  The scalar has a non-minimal coupling to
gravity determined by the dimensionless parameter~$\xi$, which allows it
to act as the inflaton, in analogy to the Higgs Inflation scenario
involving the SM Higgs \cite{Bezrukov:2007ep}.  Thus, the Lagrangian in the Jordan frame is
\begin{equation} \label{eq:u1jordanlag}
	\mathcal{L}_J = \sqrt{-g_J} \left[
	 \frac{\Mpbar^2}{2} \, \Omega^2(\phi) \, R_J
	 + g_J^{\mu\nu} (D_\mu\phi)^\dagger (D_\nu\phi)
	 - \frac{1}{4} g_J^{\mu\rho} g_J^{\nu\sigma} X_{\mu\nu} X_{\rho\sigma}
	 - V_J(\phi)
	\right] ,
\end{equation}
where $\Mpbar=\frac{1}{\sqrt{8\pi G}} \simeq 2.4 \yj{\times} 10^{18}$\,GeV is
the reduced Planck mass,
\begin{equation} \label{eq:Omega}
	\Omega^2(\phi) = 1 + 2 \xi \, \frac{\phi^\dagger\phi}{\Mpbar^2}
\end{equation}
is the conformal factor representing the non-minimal coupling,
and $R_J$ is the Ricci scalar in the Jordan frame.
The remaining quantities are defined as
\begin{align}
	D^\mu\phi &= \left( \partial^\mu + i g X^\mu \right) \phi \,, \\
	X^{\mu\nu} &= \partial^\mu X^\nu - \partial^\nu X^\mu \,, \\ 
	V_J(\phi) &= -\mu^2 \phi^\dagger\phi + \lambda \, (\phi^\dagger\phi)^2 =
	 \lambda\left(\phi^\dagger\phi - \frac{v^2}{2}\right)^2 + \text{const.},
\end{align}
where
for convenience we have chosen the $U(1)_X$ charge equal to one for the scalar.
We assume $\mu^2>0$, such that the $U(1)_X$ symmetry is broken
spontaneously by a vacuum expectation value $v= \mu/\sqrt{\lambda} \ll\Mp$,
which yields a mass $m_X = g v$ for the gauge boson at tree level.

In order to study the impact of fermionic degrees of
freedom in the FOPT analysis, we also consider the
option of adding $n_\psi$ pairs of chiral fermions ${\psi_L}_i$ and
${\psi_R}_i$ with equal $U(1)$ charges of $\frac{1}{2}$, which ensures
anomaly cancellation \cite{Baek:2023cfy}
and allows Yukawa couplings to the scalar,
\begin{equation} \label{eq:Lpsi}
	\mathcal{L}_\psi = \sqrt{-g_J} \, \sum_{i=1}^{n_\psi} \left[
	 {\overline{\psi}_L}_i i \slashed{D} {\psi_L}_i +
	 {\overline{\psi}_R}_i i \slashed{D} {\psi_R}_i -
	 \frac{y}{2} \left(
	  {\overline{\psi}_L}_i \, \phi \, {\psi_L}_i^c +
	  {\overline{\psi}_R}_i \, \phi \, {\psi_R}_i^c + \text{h.c.} \right)
	\right] ,
\end{equation}
where
\begin{equation}
	\slashed{D} {\psi_{L,R}}_i =
	\gamma^m e^\mu_m \left( \partial_\mu + i \, \frac{g}{2} X_\mu + \omega_\mu \right) {\psi_{L,R}}_i
\end{equation}
and $\omega_\mu(x)$ is the spin connection.%
\footnote{Here, we used the Latin index $m$ for the local inertial
coordinates $\{\xi^m\}$ and the Greek index $\mu$ for the general coordinates $\{x^\mu\}$. Then $\sqrt{-g_J}$ corresponds to the determinant of the vierbein $e^m_\mu=d\xi^m/dx^\mu$.
}
The superscript ${}^c$ denotes charge conjugation.
We assume a universal Yukawa coupling~$y$ for simplicity.
We also assume the potential mass terms
$-m {\overline{\psi}_L}_i {\psi_R}_i$ to be irrelevant, i.e., $m \ll v$.
After spontaneous symmetry breaking, we obtain $2 \, n_\psi$ Majorana fermions
${\psi_L}_i + {\psi_L}_i^c$ and ${\psi_R}_i + {\psi_R}_i^c$ with mass
$m_f = \frac{y}{\sqrt{2}} v$ at tree level.

We have included the coupling to gravity of all particles, but only the gravitational interactions of the scalar will be relevant in the discussion of inflation in the following.

The purpose of our studies is to determine the conditions under which a
strong FOPT that can produce observable GW can take place, and hence we
want to explore scales between the Planck mass and much lower energies. For
this purpose, we need to take into account the renormalization group (RG)
evolution of the couplings in the Lagrangians \eqref{eq:u1jordanlag} and
\eqref{eq:Lpsi} between $\Mpbar$ and the symmetry-breaking scale $v$ at
which a FOPT can occur.  Using SARAH~\cite{Staub:2008uz,Staub:2013tta},
we obtain the 1-loop RG equations
\begin{align}
    \label{eq:RGE_g}
	16\pi^2 \, \mu_R \frac{dg}{d\mu_R} &= \frac{n_\psi+1}{3} g^3 \,,
\\
    \label{eq:RGE_y}
	16\pi^2 \, \mu_R \frac{dy}{d\mu_R} &= (n_\psi+1) \, y^3 - \frac{3}{2} g^2 y \,,
\\
	\label{eq:BetaLambda}
	16\pi^2 \, \mu_R \frac{d\lambda}{d\mu_R} &= 
	20 \lambda^2 - 12 g^2 \lambda + 4 n_\psi \, y^2 \lambda + 6 g^4 - 2 n_\psi \, y^4
\end{align}
above the symmetry-breaking scale ($\mu_R>v$),
where $\mu_R$ is the renormalization scale.

For $\lambda, y \ll g$, the evolution of the scalar self-coupling is
dominated by the gauge interactions, which decrease its value when
running from high to low energies.  As we have to require a positive
value at the symmetry-breaking scale, this leads to a lower bound on
$\lambda$ at the Planck scale.  Conversely, given a value of $\lambda$
yields an upper limit on $g$.  Approximating the r.h.s.\ of
eq.~\eqref{eq:BetaLambda} to be constant and dominated by $6g^4$, we
estimate the running to change $\lambda$ by
\begin{equation}
	\lambda(v) - \lambda(\Mpbar) \simeq
	-\frac{6 g^4}{16\pi^2} \ln\frac{\Mpbar}{v} \simeq
	-0.6 \, g^4 \,.
\end{equation}
Consequently, we need
\begin{equation}
	\lambda(\Mpbar) \gtrsim 0.6 \, g^4 \,,
\end{equation}
where the scale at which $g$ is evaluated does not matter much, since
$g$ runs fairly slowly.

If the Yukawa coupling dominates the RG evolution, $\lambda, g \ll y$, the
relevant term on the r.h.s.\ of eq.~\eqref{eq:BetaLambda} is $-2n_\psi\,y^4$,
so $\lambda$ increases towards smaller energies, and
\begin{equation}
	\lambda(v) - \lambda(\Mpbar) \simeq
	\frac{2 n_\psi \, y^4}{16\pi^2} \ln\frac{\Mpbar}{v} \simeq
	0.2 \, n_\psi \, y^4 \,.
\end{equation}

For large values of the couplings, the running is no longer linear and
the above approximations become invalid.  Some couplings can become too
large for perturbative calculations to be reliable.  This is most likely to
happen for the scalar self-coupling.  For example, for $n_\psi=1$, $v = 10^{12}$\,GeV,
and $\lambda, y \ll g$ at that scale, the gauge coupling must be smaller
than about $1.4$ to ensure a sufficiently small value of $\lambda$ up to
the Planck scale.

If $g$ and $y$ are of similar size, it is possible to combine their
effects on the RG evolution to realize small values of $\lambda$ at both $v$
(as required for a FOPT, see Sec.~\ref{sec:BenchmarkFOPT}) and $\Mpbar$
(as required for Critical Higgs Inflation, see Sec.~\ref{sec:InflationObs}).
Alternatively, we could add fermions with vectorlike masses larger than
$v$ that modify the running but not the PT\@.
Of course, both options require fine-tuning.

We assume a coupling between the dark sector and the SM
particles (visible sector) that is sufficiently strong to keep both
sectors in thermal equilibrium with a common temperature but
sufficiently weak to justify neglecting the impact of SM particles on
the dark Higgs effective potential and the phase transition.
The latter also ensures that collider searches do not constrain the dark
sector for $v \gtrsim 10^4$\,GeV\@.
The simplest options for such a coupling are a Higgs portal as well as a
kinetic mixing term between the dark gauge boson and the $U(1)_Y$ gauge
boson.
After the phase transition, this coupling allows the remaining dark sector particles to decay to SM particles.%
\footnote{If one of the dark sector particles is stable, it is a dark matter candidate.  We leave an exploration of this option for future work.}
Hence, the cosmological evolution at temperatures sufficiently below the $U(1)_X$ symmetry breaking scale is unchanged compared to the SM.

\subsection{Scalar potential}
\subsubsection{Effective potential at finite temperature}
We write the complex scalar in terms of two real fields,
\begin{equation}
	\phi = \frac{1}{\sqrt{2}} \left( h + i \chi \right) .
\end{equation}
By gauge invariance, the effective potential depends only on
\[
	\phi_c^\dagger\phi_c = \frac{1}{2} \left( h_c^2 + \chi_c^2 \right) ,
\]
where the subscript $c$ denotes a classical field.  Consequently, we can
set $\chi_c=0$ without loss of generality and write the effective potential as a function
of $h_c$. Then the effective potential at finite temperature for the dark sector is given by
\begin{equation}
V(h_c,T)=V_\text{tree}(h_c)\, +
\sum_{i=h,f,g,\chi}\big(V_{i,1l}(h_c)+V_{i,\rm{th}}(h_c,T)\big) \,,
\label{eq:completev1lu1}
\end{equation}
where the indices $h,f,g,\chi$ refer to the scalar, fermion, gauge boson
and Nambu-Goldstone boson contributions, respectively. The tree-level potential is given by
\begin{equation}
    V_{\text{tree}}(h_c)=-\frac{\mu^2}{2}h_c^2+\frac{\lambda}{4}h_c^4 \,.
\end{equation}
The function
$V_{i,1l}(h_c)$ is the 1-loop correction at zero temperature
\cite{Coleman:1973jx} and $V_{i,\text{th}}(h_c,T)$ is the 1-loop
correction at finite temperature \cite{PhysRevD.9.3320,Linde:1981zj},
are given for completeness in \eq{eq:effectivecorrpot} of App.~\ref{app:thermal}. These
functions depend on the field-dependent masses
\begin{eqnarray}
m_h^2(h_c)&=&-\mu^2+3\lambda h_c^2 \,,
\nonumber\\
m_\chi^2(h_c) &=& -\mu^2+\lambda h_c^2 \,,
\nonumber\\
m_g^2(h_c) &=& g^2 h_c^2 \,,
\nonumber\\
m_f^2(h_c) &=& \frac{y^2}{2}h_c^2 \,.
\label{eq:fielddepmu1}
\end{eqnarray}
The 1-loop functions also depend on the bosonic and fermionic degrees of
freedom
\begin{equation}
\label{eq:dofu1dark}
n_g=3 \,,\quad n_h=1 \,,\quad n_\chi=1 \,,\quad n_f = -4 n_\psi \,,
\end{equation}
where the latter are defined as a negative number for convenience.
Of course, $n_\psi=0$ for the case without fermions.

\subsubsection{Effective potential for Dark Higgs Inflation \label{subsubsec:EffpotEf}}
The calculation of the scalar potential during the inflationary phase
proceeds in analogy to the well-known Higgs Inflation scenario
\cite{Bezrukov:2007ep,Bezrukov:2013fka}.  We include it here for clarity
and to fix the notation.
For brevity, we set $\Mpbar=1$ in this section.
Starting from the Jordan frame Lagrangian~\eqref{eq:u1jordanlag}, the
Lagrangian in the Einstein frame can be obtained by the Weyl transformation
\begin{align}
g_{J\mu\nu} &\to g_{\mu\nu} \equiv \Omega^2 g_{J\mu\nu} \,,\\
g_J^{\mu\nu} &\to g^{\mu\nu} \equiv \Omega^{-2} g_J^{\mu\nu} \,,\\
g_J ={\rm det}(g_{J\mu\nu}) &\to g = \Omega^8 g_J \,,\\
R_J =g_J^{\mu\nu}R_{J\mu\nu} &\to R =
 \Omega^{-2} R_J + \frac{3}{2}g^{\mu\nu}(\partial_\mu \ln\Omega^2)(\partial_\nu \ln\Omega^2) - 3\, \square \ln\Omega^2 \,,
 \label{eq:RJ}
\end{align}
where the subscript $J$ refers to quantities in the Jordan frame,
and quantities without this subscript are understood to be in the
Einstein frame.
The last term in \eq{eq:RJ} is a total divergence and can be dropped.
Using unitary gauge, $\phi=\frac{1}{\sqrt{2}}h$,
and switching to the canonically normalized field $\zeta$,
the scalar Lagrangian becomes
\begin{equation}
\cL_\zeta = 
\sqrt{-g}\left[
\half R+\half(\partial_\mu\zeta)(\partial^\mu\zeta)-V(\zeta)+
(\text{gauge interactions})\right],
\end{equation}
where we have identified the potential in the Einstein frame as
\begin{equation} \label{eq:VE}
V(\zeta) = \frac{V_J\big(h(\zeta)\big)}{\Omega^4\big(h(\zeta)\big)} =
\frac{\frac{\lambda}{4}(h^2-v^2)^2}{\parent{1+\xi h^2}^2}
\end{equation}
and where the canonically normalized scalar is obtained via
\begin{equation} \label{eq:Fhdef}
\zeta = \int_0^h dh' \, \sqrt{\frac{6\xi^2 {h^\prime}^2}{\Omega^4(h')}+\frac{1}{\Omega^2(h')}}
\,.
\end{equation}
The integration can be performed in a straightforward way,
\bea
\zeta &=&
\sqrt{\frac{1}{\xi }+6} \sinh^{-1}\left(h \sqrt{\xi (6 \xi +1)}\right)
-\sqrt{6} \tanh^{-1}\left(\frac{\sqrt{6} h \xi }{\sqrt{h^2 \xi (6 \xi +1)+1}}\right)
\nn\\
&\simeq&
\begin{cases}
\begin{alignedat}{3}
    &h&\,, \quad& h\ll\frac{1}{\sqrt{\xi}} \\
    &\sqrt{6+\frac{1}{\xi}}\ln(h\sqrt{\xi})& \,, \quad& h\gg\frac{1}{\sqrt{\xi}}
\,.
\end{alignedat}
\end{cases}
\eea
Then the potential in the Einstein frame can be expressed in the two regimes as
\be
\label{eq:asymptotic_VE}
V(\zeta)\simeq
\begin{cases}
\begin{alignedat}{3}
    &\frac{\lambda}{4}\zeta^4& \,, \quad& v\ll h\sqrt{\xi}\ll 1 \\
    &\frac{\lambda}{4\xi^2}\parent{1+\exp\parent{-\frac{2\zeta}{\sqrt{6+1/\xi}}}}^{-2}& \,, \quad & h\sqrt{\xi}\gg 1 \,.
\end{alignedat}
\end{cases}
\ee
Consequently, the potential becomes flat for large field values, which
enables slow-roll inflation.

\section{Parameter space compatible with inflation and FOPT \label{sec:foptresults}}

\subsection{First-order phase transitions \label{sec:BenchmarkFOPT}}

Our final goal is to identify concrete conditions for the parameter
space where we can achieve a strong FOPT in general dark sector gauge
models, but we leave this discussion for later work
\cite{sinchong2:2024}. As a proof of principle, we present here several
benchmark points for which a FOPT leading to measurable GW can occur.

FOPT can proceed via different processes: nucleation, expansion,
collision and merger of bubbles of the broken phase. The way these
processes can source GW is quite varied. If the phase transition occurs
only in vacuum, the only source of GW is the collision of the bubble walls and the GW density as a function of the frequency can be expressed in terms of the envelope approximation \cite{Huber:2008hg}. If the scalar bubbles expand instead in a hot plasma, the plasma friction will slow down the expansion of the walls and so most of the energy released by the transition will be transferred from the scalar field into the plasma, making the contribution to GW from the scalar field subdominant in comparison to that of the plasma. Numerical simulations of this process on a 3D lattice have shown that the energy-momentum tensor of the fluid after bubble collisions corresponds to an ensemble of sound waves \cite{Hindmarsh:2013xza,PhysRevD.92.123009} and that this process dominates over all over the others in the case of the hot plasma. This dominance is the so-called acoustic phase. A competing source of GW is turbulence \cite{Kosowsky:2001xp,Nicolis:2003tg,Caprini:2006jb,PhysRevD.86.103005,Kisslinger:2015hua} created in the plasma, and in fact for very strong transitions the acoustic phase turns over into a turbulent stage which continues to produce gravitational radiation until it decays. In order to determine the dependence of the GW on the parameters of the phase transition and on the frequency it is necessary to capture the shock waves in the plasma and for this numerical simulations are needed. We use up-to-date GW templates (fits in terms of frequency power laws and parameters of the theory) from 3D lattice simulations 
\cite{Hindmarsh:2013xza,Hindmarsh:2015qta,Caprini:2009yp,Binetruy:2012ze,Ellis:2018mja}. Simulations and fits of the GW density are an active line of research and continuously improving. The useful information is to locate the peak frequency and to establish an order of magnitude of the GW density.  One crucial ingredient of the FOPT is the parameter $\alpha$, which is the ratio of the vacuum energy density released in the transition and of the plasma energy density in the symmetric phase. For our case,
\begin{equation}
\alpha=\frac{1}{\rho_{\rm rad}}\left.\left[\Delta V(h_c,T)-T\frac{dV(h_c,T)}{dT}\right]\right|_{T=T_n} ,
\label{eq:defalpha}
\end{equation}
which turns out to be the ratio of the latent heat to the radiation
density in the case of GW production during the radiation-dominated era.%
\footnote{The observed upper limit on the effective number of neutrino species
$N_\text{eff}$ places an upper limit on $\alpha$.} 
Here $\Delta V(h_c,T)$ is the difference of the effective potential of \eq{eq:completev1lu1} between the true minimum and the false minimum and $T_n$ is the nucleation temperature.
The derivative is evaluated at the true potential minimum.

For our study, we restrict ourselves to benchmark
points with $\alpha \sim 0.1$, a value which is safe since the radiation density increases relative to the GW density as the Universe
cools and particles annihilate.
Furthermore, the lattice simulations on which the templates that we use are valid only in the $\alpha \lesssim 0.1$ regime \cite{Hindmarsh:2013xza,Hindmarsh:2015qta,Caprini:2009yp,Binetruy:2012ze}.\footnote{Recent templates have been obtained in \cite{Jinno:2022mie} but for weak ($\alpha=0.0046$) and intermediate ($\alpha=0.05$) FOPT in a hybrid 1D lattice-analytical approach. We comment on this in App.~\ref{app:GWfor}.}
For completeness, we include the formulas we use in App.~\ref{app:GWfor}\@.
Gravitational wave frequency density templates for $\alpha>1$ exist but they require more knowledge about the difference between the reheating temperature and the nucleation temperature. For the case of small $\alpha$ and the barrier of the potential produced mainly by thermal effects, the reheating temperature and the nucleation temperature can be safely identified with each other. 
Another crucial parameter in the computation of GW density from FOPT is the beta parameter 
\bea
\label{eq:betadef}
\beta=H_*\left.T\frac{d}{dT}\left(\frac{S_3}{T} \right)\right|_{T=T_n\approx T_*}
,
\eea
which quantifies how fast or slow the FOPT occurs. Here $T_*$ is the temperature of the thermal bath and  $S_3$ is the 3D action
\begin{equation}
\label{eq:S3T}
S_3[h_c(r),T] = 4 \pi \int_0^\infty r^2 dr \left[ \frac{1}{2} \left( \frac{dh_c(r)}{dr} \right)^2 + V(h_c(r),T) \right],
\end{equation}
where $r$ represents the radius of the bubble and $h_c(r)$ the instanton solution.

\subsubsection{Benchmark points without fermions}
We are interested in finding values of $g$ and $\lambda$ for which observable GW spectra can be produced as a result of a FOPT\@.
In the case of no fermions,
we find a FOPT leading to acceptable GW parameters for $g(v)=0.95$ and
$\lambda(v)=10^{-3}$, where $v$ is close to the scale at which the FOPT occurs.
In \Tabref{table:model_params} we present four benchmark points (BP) for this case, which differ in the value of $v$. 
We also show the characteristic temperatures of the FOPT, $T_0$, $T_c$ and $T_n$, which are, respectively, the temperature at which $\partial^2 V(h_c,T)/\partial h_c^2=0$ at the origin, the critical temperature (at which the two minima are degenerate) and the nucleation temperature (the temperature at which one bubble is nucleated per Hubble horizon on average, see \eq{eq:NTnO1}). Finally, the parameters $\alpha$ and $\beta/H_*$, defined in Eqs.~\eqref{eq:defalpha} and \eqref{eq:betadef}, are given.
We refer to App.~\ref{app:FOPTdet} for details of the computation.

\begin{table}[bt]
    \centering
    \begin{tabular}{|l|c|c|c||c|c|c|c|c|c|}
    \hline
    \hline
    BP & $g(v)$ & $\lambda(v)$ &   $v/\rm{GeV}$ & $T_0/v$ & $T_c/v$ &  $T_n/v$ & $\alpha$ &  $\beta/H_*$\\
    \hline
    \hline
    BP1 & 0.95 & $10^{-3}$ &  $10^{12}$ &0.068 & 0.37 &  0.11 & 0.94 & $1.3\times 10^2$\\
    BP2 &  &  & $10^{9}$ &      &  &0.13 & 0.45 & $1.5\times 10^2$ \\
    BP3 &  &  &$10^{6}$   &     &  &  0.16 & 0.25 & $1.9\times 10^2$\\    
      BP4 &  &  & $10^{3}$ &    &  & 0.18 & 0.16 & $2.4\times 10^2$\\
               \hline
    \end{tabular}
    \caption{Benchmark points leading to a FOPT for the case without fermions ($n_\psi=0$).
    }
    \label{table:model_params}
\end{table}

The way the 1-loop corrections contribute to the effective potential \eqref{eq:completev1lu1}, at zero temperature and at finite temperature, are shown in \Figref{fig:Potcomparisons}. Given the small size of $\lambda$, the tree-level contribution is subdominant and the 1-loop contributions and the thermal contributions are necessary for the formation of the potential barrier,\footnote{Note that in this case the fact that 1-loop contributions can overcome the tree level is natural and we are in a regime where 1-loop contributions are reliable since $\lambda$ is chosen to be small.}
as can be seen in the bottom left plot of \Figref{fig:Potcomparisons}. The expansion of $J_b(m^2_i(h_c)/T^2)$ in terms of powers and logarithms of $m^2_i(h_c)/T$ given in \eq{eq:effectivecorrpot} is the high-temperature approximation and even if it does not match entirely the shape of the complete thermal corrections (see top and middle plots of \Figref{fig:Potcomparisons}), it is useful to understand the behavior of the barrier. In this case, both the cubic and logarithmic contributions are important to produce a barrier. As these contributions are proportional to $g^3$ and $g^4$, respectively, both the barrier height and $\alpha$ increase significantly with increasing gauge coupling.
In \Tabref{table:model_params} we show the FOPT parameters that we find for our choice of parameters ($\lambda$ and $g$ for different scales: 
$v/\text{GeV} \in \{ 10^3, 10^6, 10^9, 10^{12} \}$).

As the quantity $S_3/T$ is dimensionless (cf.\ \eq{eq:S3T}), we can
compute everything related to the phase transition in terms of
dimensionless quantities, except for $T_n$, which depends on the
probability of nucleation, represented by \eq{eq:NTnO1}.  Note the
factor $T^5$ in the denominator of the integral in \eq{eq:NTnO1}; when
the symmetry-breaking scale $v$ increases, the $T^{-5}$ suppression
becomes more severe and makes it more difficult for the integral between
$T_n$ and $T_c$ to reach an $\mathcal{O}(1)$ value, which is required
for completing the phase transition.  Thus, increasing $v$ causes a
decrease of $T_n/v$; if $v$ becomes too large, $T_n$ approaches $T_0$
(where the potential barrier disappears) and a FOPT is not feasible.

\begin{figure}
    \includegraphics[width=1.0\linewidth]{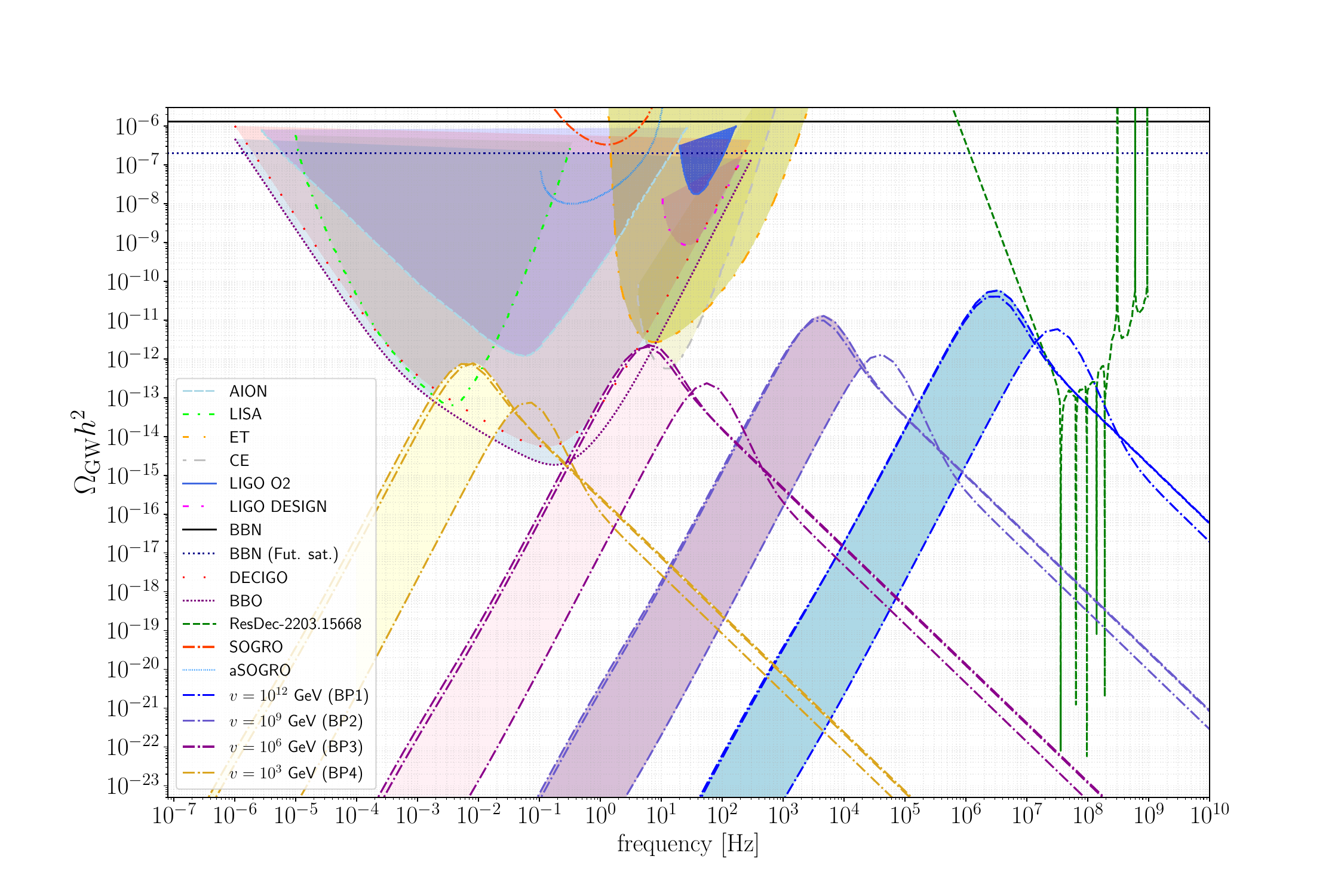}
    \caption{
    Gravitational wave spectra for the benchmark points BP1 to BP4 (from right to left), which are defined in \Tabref{table:model_params} and correspond to $v=(10^{12}, 10^{9}, 10^{6}, 10^{3})$\,GeV, respectively. We remind the reader that the formulas for GW have been tested in a regime $\alpha\lesssim 0.1$. The value of $\alpha$ obtained for BP1 ($v=10^{12}$\,GeV) is close to 1, so the result must be taken with caution. For every case we have plotted three bubble wall velocities, corresponding, from left to right, to $v_w=1$, $v_w=v_w^d$ (the detonation velocity) and $v_w=0.1$. We note that for some of the cases the first two lines are indistinguishable due to the proximity of $v_w^d$ to 1.}
    \label{fig:GW_N1_density}
\end{figure}

In \Figref{fig:GW_N1_density} we show the GW densities we obtain for the benchmark points of \Tabref{table:model_params} 
{by summing the contributions from sound waves and
turbulence given in \eq{eq:Omegasw} and \eq{eq:Omegaturb}, respectively.}
We plot each of the benchmark points with three different bubble wall velocities, $v_w$, since this quantity dominates the uncertainty in the determination of the GW spectrum, as we explain in App.~\ref{app:FOPTdet} and \ref{app:GWfor}\@.

Most importantly, the choice of the symmetry-breaking scale determines
the peak frequency of the GW spectrum.
The peak height decreases with decreasing $v$, reflecting the decrease
of $\alpha$.
The shape of the spectrum is very similar for all benchmark points.

The figure also displays the current
(for the case of LIGO \cite{LIGOScientific:2019vic}) and expected sensitivities for different experiments: AION \cite{Badurina:2019hst}, LISA \cite{2017arXiv170200786A}, Einstein Telescope (ET) \cite{Maggiore:2019uih}, Cosmic Explorer (CE) \cite{Evans:2021gyd}, DECIGO \cite{Kawamura:2020pcg}, BBO \cite{BBOref}, SOGRO \cite{10.1093_ptae045} and a proposed resonant detector \cite{Herman:2022fau} operating in the ($10^6$, $10^9$)\,Hz frequency range. SOGRO is a superconducting tensor detector for mid-frequency GW proposed to fill the gap between LISA and LIGO\@.
{In addition, we include the current Big Bang Nucleosynthesis (BBN) bound as well as a forecasted \cite{Pagano:2015hma} improved bound, which is achievable with future satellite experiments such as COrE \cite{thecorecollaboration2011core} and EUCLID \cite{laureijs2011euclid}.}

As is well known, GW from FOPT around $v=10^3$\,GeV are expected to peak in the sensitivity range of LISA, as we see from the GW density on the left of the plot in \Figref{fig:GW_N1_density}.
If $v=10^3\,$GeV, the
dark sector can be constrained by LHC searches, in particular invisible
Higgs decays into the Dark Higgs, whose mass is about $45\,$GeV in this
case.  The relevant parameter is the quartic coupling between the SM
Higgs and the Dark Higgs (Higgs portal), whose value can be adjusted
rather freely in our scenario.  For fairly large kinetic mixing
between the $X^\mu$ and the SM $U(1)_Y$ gauge boson, the new vector state
can also be accessible.

For the case of $v=10^6$\,GeV, CE has the potential to observe the
signal. For $v=10^9$\,GeV, future ground-based detectors (not shown in
the figure) can probe the relevant area, in principle. For the case of
$v=10^{12}$\,GeV, resonant detectors appear promising, which have recently been contemplated as a way to probe the so-called \emph{ultra-high-frequency} region (UHF)\@. We can see that our BP1 lies precisely in the resonant region of the detector proposed in \cite{Herman:2022fau}. 

\begin{figure}
    \centering
    \includegraphics[width=0.8\textwidth]{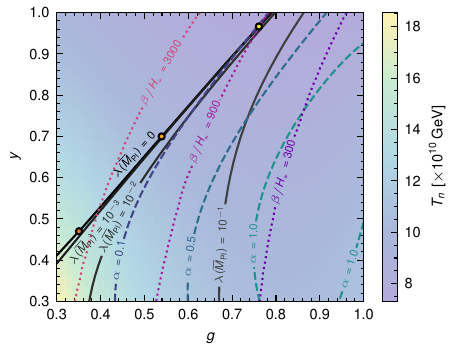}
    \caption{Parameter space in the ($g$--$y$) plane at the scale of $10^{12}$\,GeV. The solid lines represent $\lambda(\Mpbar)$ contours for values $\{0, 10^{-3}, 10^{-2}, 10^{-1}\}$, while the dotted and dashed lines show $\beta/H_*$ contours ($3000, 900, 300$) and $\alpha$ contours ($0.1, 0.5, 1.0$), respectively.  The benchmark points BP5--BP7 defined in \Tabref{table:params_wf} are denoted by circle markers. The background color gradient represents the nucleation temperature $T_n$ in units of $10^{10}$\,GeV.}
    \label{fig:grid_search_wf}
\end{figure}

\subsubsection{Benchmark points with fermions}
Now we turn to the case with fermions. As we can see from \eq{eq:effectivecorrpot} and the expansion of $J_f$ in the regime of $m_i^2(h_c)/T^2 \ll 1$, the fermion contribution to the effective potential does not contain a cubic term and therefore fermions do not enhance the barrier. In \Figref{fig:Potcomparisons} we show the exact thermal behavior along with different approximations and contributions. As opposed to the case of no fermions, the high-temperature approximation is not a good approximation for the complete duration of the phase transition, as the temperature corrections are led mainly by the logarithmic term and to some extent the linear term. 

Combining the effects of gauge bosons and fermions on the RG evolution of the scalar self-coupling, it is possible to arrive at $\lambda=0$ at the Planck scale.
In \Tabref{table:params_wf} we show three different benchmark points
realizing this possibility for three different choices of gauge and
Yukawa couplings at the symmetry-breaking scale $v=10^{12}$\,GeV\@.
{Of course, the initial conditions have to be tuned with
a higher precision than given in the table.  Besides, the exact location
where $\lambda$ reaches zero is sensitive to higher loop orders.
However, these complications do not change the fact that
$\lambda(\Mpbar)=0$ is possible \emph{in principle} and have no impact
on the FOPT and the resulting GW.}

\begin{table}[bt]
    \centering
    \begin{tabular}{|l|c|c|c|c||c|c|c|c|c|}
    \hline
    \hline
    BP & $g(v)$ & $\lambda(v)$ &  $y(v)$  & $v/\rm{GeV}$ & $T_0/v$ & $T_c/v$ & $T_n/v$ & $\alpha$ &  $\beta/H_*$\\  
    \hline
    \hline
    BP5 & 0.76 & $10^{-3}$ &  0.97 & $10^{12}$ & 0.067 & 0.11 & 0.075 & 0.10 & $9.1\times 10^{2}$ \\
    BP6 & 0.54 & $10^{-3}$ &  0.70 & $10^{12}$ & 0.094 & 0.13 & 0.10 & 0.070 & $1.7\times 10^{3}$ \\
    BP7 & 0.35 & $10^{-3}$ &  0.47 & $10^{12}$ & 0.14 & 0.16 & 0.15 & 0.024 & $3.9\times 10^{3}$\\
    \hline
    \end{tabular}
    \caption{Benchmark points with $n_\psi=1$ and $\lambda(\Mpbar) \approx 0$, which are shown in \Figref{fig:grid_search_wf}.
    }
    \label{table:params_wf}
\end{table}

\begin{figure}
    \includegraphics[width=1.0\linewidth]{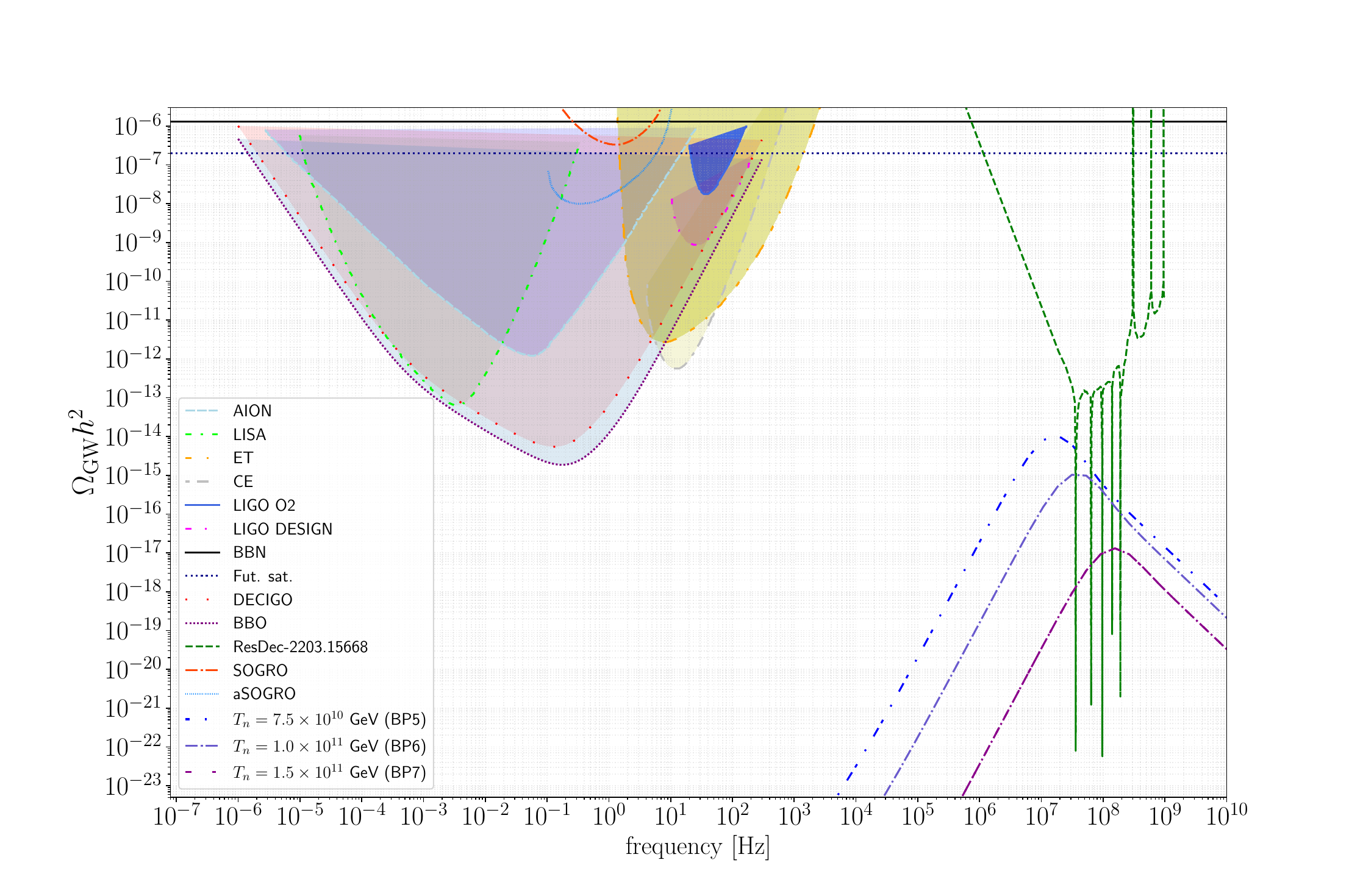}
    \caption{Gravitational wave spectra for the cases of \Tabref{table:params_wf}, that is, including a fermion pair. We choose three benchmark points (BP5--BP7) along the line
$\lambda(\Mpbar)$=0 of \Figref{fig:grid_search_wf}. For all these cases, $v=10^{12}$\,GeV but due to the different gauge couplings, the resulting nucleation temperature $T_n$ is different. 
    \label{fig:GW_density_withf}}
\end{figure}

The corresponding GW densities are shown in \Figref{fig:GW_density_withf}, along with the reach of planned and proposed experiments. Here we plot only the GW density profile that corresponds to the detonation bubble wall velocity $v_w^d$ (see App.~\ref{app:GWfor}) in order not to overcrowd the region about the density profiles. As the values of the Yukawa and gauge couplings are different, the corresponding nucleation temperatures are different. Interestingly, the GW density profiles for these cases peak in the region of sensitivity of the resonant detector proposed in \cite{Herman:2022fau}. Their height is smaller than in the case without fermions since the gauge coupling is smaller.

We could in principle look for benchmark points at lower scales than $v=10^{12}$\,GeV, as in the case with fermions. However, since the lower the scale the more difficult it is to obtain a large peak amplitude, we do not present them. The qualitative behavior is the same as without fermions, i.e., the peak frequency decreases for decreasing $v$. The main importance of the case with fermions is that we can drive to zero the value of $\lambda$ at $\Mpbar$.

\subsection{Inflationary observables} \label{sec:InflationObs}

As discussed in Sec.~\ref{subsubsec:EffpotEf},
the scalar potential in the Einstein frame, given in Eq.~\eqref{eq:asymptotic_VE}, has an asymptotic plateau
in the large-field limit.  Putting back the Planck mass,
\begin{align}
V(\zeta) &\approx \frac{\lambda \Mpbar^4}{4\xi^2}\parent{1+\exp\parent{-\frac{2\zeta/\Mpbar}{\sqrt{6+1/\xi}}}}^{-2} 
\approx \frac{\lambda \Mpbar^4}{4\xi^2} \;, \quad \zeta \gg \Mpbar/\sqrt{6+1/\xi} \,.
\end{align}
This plateau is responsible for the slow-roll inflationary dynamics, which enables the exponential expansion of the Universe. The slow-roll parameters are expressed in terms of the Einstein-frame potential and the canonically normalized field $\zeta$,
\bea
    \varepsilon&=&\frac{\Mpbar^2}{2}\parent{\frac{dV/d\zeta}{V}}^2\approx\frac{8\xi}{1+6\xi}\exp\parent{-\frac{4\zeta/\Mpbar}{\sqrt{6+1/\xi}}},\\
    \eta&=&\Mpbar^2 \frac{d^2 V/d\zeta^2}{V}\approx-\frac{8\xi}{1+6\xi}\exp\parent{-\frac{2\zeta/\Mpbar}{\sqrt{6+1/\xi}}}.
\eea

In the slow-roll framework, we can relate the inflationary observables
with the slow-roll parameters and the potential at the pivot scale
$\zeta_*$ setting
$N_*=\int_{\zeta_e}^{\zeta_*}d\zeta/(\Mpbar\sqrt{2\varepsilon})=60$
with $\max(\varepsilon,|\eta|)_{\zeta_e}=1$ at the end of inflation. The
observables are conveniently related to the number of $e$-foldings as
\begin{align}
n_s&=1-6\varepsilon_*+2\eta_* \approx 1-\frac{2}{N_*} -\frac{9}{2N_*^2}\approx 0.965 \,, \\
r&=16\varepsilon_* \approx \frac{12}{N_*^2} \approx 0.003 \,,
\end{align}
where we took $\zeta_*\ll \zeta_e$ and neglected the correction $1/(6\xi N_*^2) \lesssim 4\times 10^{-4}$ for $\xi\gtrsim 1$. 
These values are in excellent agreement with the Planck2018 results \cite{Planck:2018jri}, $n_s=0.965\pm 0.004$ (for $k_*=0.05\,\text{Mpc}^{-1}$) and $r<0.035$ (95\% CL),
as in conventional Higgs Inflation models.
The CMB normalization provides a condition for $\lambda/\xi^2$,
\begin{align}
    A_s=\frac{1}{24\pi^2}\frac{V_*}{\varepsilon_*} \approx \frac{\lambda(\Mpbar) \, N_*^2}{72\pi^2 \xi^2}.
\end{align}
Using $A_s=(2.098\pm0.023)\times 10^{-9}$ \cite{Planck:2018jri}, we conclude
\begin{align} \label{eq:lambdaxi}
    \frac{\lambda(\Mpbar)}{\xi^2} \approx 3.9\times 10^{-10}
\end{align}
for $N_*=60$ $e$-foldings. 

We show the resulting values of $\xi$ for our seven benchmark points in \Tabref{tbl:pars_at_Mpbar}, together with the values of the other couplings at $\Mpbar$.

\begin{table}
    \centering
    \begin{tabular}{|l|c|c|c|c|c|}
    \hline
    \hline
    BP & $g(\Mpbar)$ & $\lambda(\Mpbar)$  & $y(\Mpbar)$ & $n_\psi$ & $\xi$\\  
    \hline
    \hline
    BP1 & 0.98 & $0.37$  & - & 0& $3.1\times 10^4$\\
    BP2 & 0.99 & $0.57$ & - & 0& $3.8\times 10^4$\\
    BP3 & 1.00 & $0.92$  & - &0  & $4.9\times 10^4$\\
    BP4 & 1.02 & $2.22$  & - & 0& $7.5\times 10^4$\\
    BP5 & $0.79$ & $\approx 0$  & $1.10$ & 1& $\mathcal{O}(1)$ \\
    BP6 & $0.55$ & $\approx 0$  & $0.74$ & 1& $\mathcal{O}(1)$ \\
    BP7 & $0.35$ & $\approx 0$  & $0.48$ & 1& $\mathcal{O}(1)$ \\
    \hline
    \end{tabular}
    \caption{Model parameters at $\Mpbar$.
BP1--BP4 correspond to the case of no fermions and BP5--BP7 to the case of $n_\psi=1$.}
    \label{tbl:pars_at_Mpbar}
\end{table}

If we insist that the naïve cut-off scale of the model, given by $\Mpbar/\xi$ or $\Mpbar/\sqrt{\xi}$~\cite{Burgess:2009ea, Barbon:2009ya, Lerner:2009na}, not lie much below the Planck scale, we are motivated to consider {the conformal scenario with} $\lambda(\Mpbar) \ll 1$,
which is analogous to the Critical Higgs Inflation scenario with the SM scalar~\cite{Hamada:2014wna,Hamada:2014iga}.
In this case, small values of $\xi$ are consistent with \eq{eq:lambdaxi}.
It is realized {at the benchmark points BP5--BP7, where} the RG running of the self-coupling $\lambda$ is affected by a large negative contribution from fermions, as discussed above.

The precise value of $\lambda$ at the pivot scale is slightly modified by the positive contribution from the gravitational interactions (see, e.g., \cite{Futamase:1997fq}).

\section{Conclusions \label{sec:conclusions}}

In this study, we explored an extension of the Standard Model incorporating a simple dark sector with a $U(1)_X$ gauge symmetry. We identified a compelling scenario where a single scalar field drives both slow-roll inflation and spontaneous symmetry breaking, leading to the production of gravitational waves (GWs). The predicted GW spectrum peaks at frequencies between $10^{-2}$ Hz and $10^8$ Hz, making it accessible to future GW observatories.

Inflation in this model is realized analogously to the Higgs inflation scenario, but with the Standard Model Higgs replaced by the Dark Higgs, which breaks the $U(1)_X$ symmetry. Notably, the model accommodates the particularly appealing case of Critical Dark Higgs Inflation, where the inflaton field’s self-coupling at the inflationary scale is fine-tuned via renormalization group running. This behavior emerges from significant fermionic contributions in loop corrections, while the bosonic contributions from gauge fields ensure a viable first-order phase transition (FOPT) at a lower energy scale, producing detectable GWs.

Our proposed scenario offers a novel pathway to probe inflation through GW observations, complementing traditional searches for GWs produced during inflation. Additionally, the dark sector naturally provides a candidate for dark matter, addressing another critical problem in astroparticle physics.
While we focused on a simple $U(1)_X$ model, this framework can be extended to larger gauge groups. Larger groups enable faster parameter evolution through renormalization group running and can create higher potential barriers, potentially leading to stronger FOPTs and enhanced GW signals. Such extensions open further opportunities for exploring the interplay between cosmology, particle physics, and gravitational wave astronomy.

\acknowledgments

We thank Vincenzo Branchina, Injun Jeong and Stefano Scopel for useful
discussions.  J.K.\ thanks \emph{Star Trek The Next Conversation} for
inspiration -- 5 stars out of 5.
The work of L.V.S.\ is supported by the National
Research Foundation of Korea (NRF) grant no.\ RS-2023-00273508.
S.C.P.\ was supported by the NRF grant funded by the Korean government (MSIT) no.\ RS-2023-00283129 and RS-2024-00340153.
J.K.\ is supported by the NRF's \emph{Brain Pool} program under grant no.\ RS-2023-00283129.

\newpage
\appendix

\section{One-loop thermally-corrected scalar potential \label{app:thermal}}
For completeness we give here the 1-loop and thermal corrections 
\cite{Coleman:1973jx,PhysRevD.9.3320,Linde:1981zj} to
the potential, which is a function of the classical scalar field $h_c$. 
 \begin{eqnarray}
 V_{i,\text{1l}}(h_c)&=&\frac{n_i}{64\pi^2}m_i^4(h_c)\parent{\ln\frac{|m_i^2(h_c)|}{v^2}-C_i},
 \nn\\
 V_{i,\text{th}}(h_c,T) &=& \left\{ 
 \begin{array}{cl}
    \frac{n_i}{2\pi^2}T^4 \, J_b(x),   &  i= h, g, \chi \\
   \frac{n_i}{2\pi^2}T^4 \, J_f(x),   &  i=f
 \end{array}
 \right., 
 \quad x\equiv \frac{m_i(h_c)}{T},
 \nn\\
 J_b(x) &=&
 \Re \int_0^\infty dy\,y^2\ln\left[1-e^{-\sqrt{y^2+x^2}}\right], 
 \nn\\
 &\ahigh&-\frac{\pi^4}{45} + \frac{\pi^2}{12}x^2-\frac{\pi}{6}x^3-\frac{1}{32}x^4\ln\frac{|x^2|}{a_B},
 \;\; |x^2|\ll 1 \,,
 \nn\\
 J_f(x) &=&
 \Re \int_0^\infty dy\,y^2\ln\left[1+e^{-\sqrt{y^2+x^2}}\right],
 \nn\\
 &\ahigh&\frac{7\pi^4}{360}-\frac{\pi^2}{24}x^2-\frac{1}{32}x^4\ln\frac{|x^2|}{a_F},
 \;\; x^2\ll 1 \,,\\
 a_B &=&223.1,~a_F=13.94 \,. \label{eq:effectivecorrpot}
 \end{eqnarray}
 The index $i$ takes the values $h,f,g,\chi$ for the scalar, fermion, gauge boson, and Nambu-Goldstone boson contributions, respectively.
 The field-dependent masses were already given in the main text in \eq{eq:fielddepmu1}. Also, $C_i=5/6$ for gauge bosons and $C_i=3/2$ for scalars and fermions. The degrees of freedom $n_g, n_h, n_\chi$ and $n_f$ have been defined in \eq{eq:dofu1dark}.

Note that for $h_c<v$, the field-dependent scalar masses can be imaginary.
In this case, $V_\text{1l}$ approximates the \emph{modified} effective
potential, whose real part determines the energy density and is thus the
relevant quantity for the phase transition
\cite{Weinberg:1987vp,Brahm:1992ek,Delaunay:2007wb}.
Accordingly, we have defined $V_{i,\text{1l}}$ and
$V_{i,\text{th}}$ as real functions.

\section{Details about FOPT computations\label{app:FOPTdet}}

After calculating the effective potential, with 1-loop corrections at zero temperature and finite-temperature corrections, the decay rate of the false vacuum, $\Gamma(T)$ has to be computed. $\Gamma(T)$ can be obtained from \cite{Coleman:1977py,Linde:1981zj,LINDE198137} 
\bea
\label{eq:gammaT}
\Gamma(T) \simeq \max \left[T^4\left(\frac{S_3}{2 \pi T}\right)^{\frac{3}{2}} \exp \left(-S_3 / T\right) \,,\, R_0^{-4} \left( S_4/2\pi\right)^2 \exp(-S_4) \right],
\eea
where $R_0$ represents the size of the nucleation bubble in four dimensions, $S_3$ and $S_4$ are, respectively, the Euclidean actions in three and four dimensions corresponding to the $\mathit{O}(3)$ and $\mathit{O}(4)$ symmetric tunneling configurations, referred to as bounce solutions.
In all cases we have studied, the first term in the brackets turns out to be the larger one.
Approaches to compute 1-loop functional determinants around spherical symmetric backgrounds, providing the next-to-leading-order correction to the bubble nucleation rates and hence removing the uncertainties in the expression of $\Gamma(T)$ above exist \cite{Ekstedt:2023sqc}. However, the purpose of our work is to establish a proof of concept of the existence of a parameter space where a non-minimally coupled inflaton to gravity leads to observable GW from a FOPT and hence we leave a more accurate computation of $\Gamma(T)$ for future work. Furthermore, the precise value of the prefactor in the nucleation rate, \eq{eq:gammaT}, is not critically important due to the dominance of the exponential term \cite{Anderson:1991zb}. Based on this reasoning, one can simplify the calculations by setting the term $(S_3/2\pi T)^{3/2}$ to $1$.

Once $\Gamma(T)$ is computed, {the nucleation temperature $T_n$ can be
determined by calculating the temperature at which the average number of bubbles nucleated per Hubble horizon is of order 1,}
\begin{align}
    N(T_n) = \left( \frac{3\Mpbar}{\pi}\right)^4\left( \frac{10}{g_{*}}\right)^2\int_{T_n}^{T_c} \frac{d T}{T^5} \left( \frac{S_3}{2 \pi T}\right)^{3/2} \exp{(-S_3/T)} \sim 1 \,,
\label{eq:NTnO1}    
\end{align}
assuming the PT to occur in the radiation-dominated era.
All results presented in Sec.~\ref{sec:BenchmarkFOPT} were obtained setting $N(T_n)=1$.
The parameter $g_*$ is the effective number of relativistic degrees of freedom
(d.o.f.). In our case, apart from the SM d.o.f.\ $g_{*,\text{SM}}=106.75$, we have $\phi$ (a complex scalar), the $U(1)$ gauge boson, and $2n_\psi$ Majorana fermions, so
\begin{equation}
g_* = g_{*,\text{SM}} + 4 + \frac{7}{2} n_\psi \,.
\end{equation}
After $T_n$ is computed, the parameters $\alpha$ and $\beta$ follow from Eqs.~\eqref{eq:defalpha} and \eqref{eq:betadef}. We have determined these parameters with our own codes based on the bounce solutions calculated by \texttt{CosmoTransitions} \cite{Wainwright:2011kj}. 

\begin{figure}
    \centering
    \includegraphics[width=0.43\linewidth]{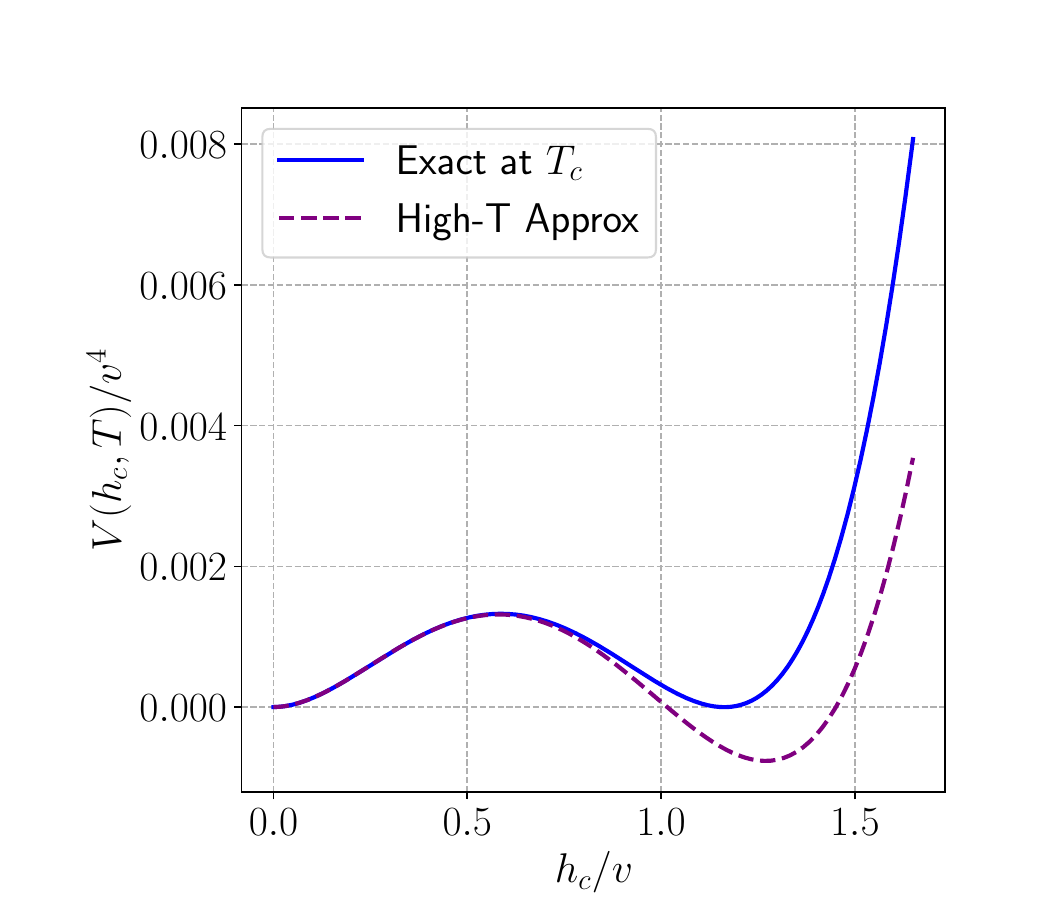}
    \includegraphics[width=0.43\linewidth]{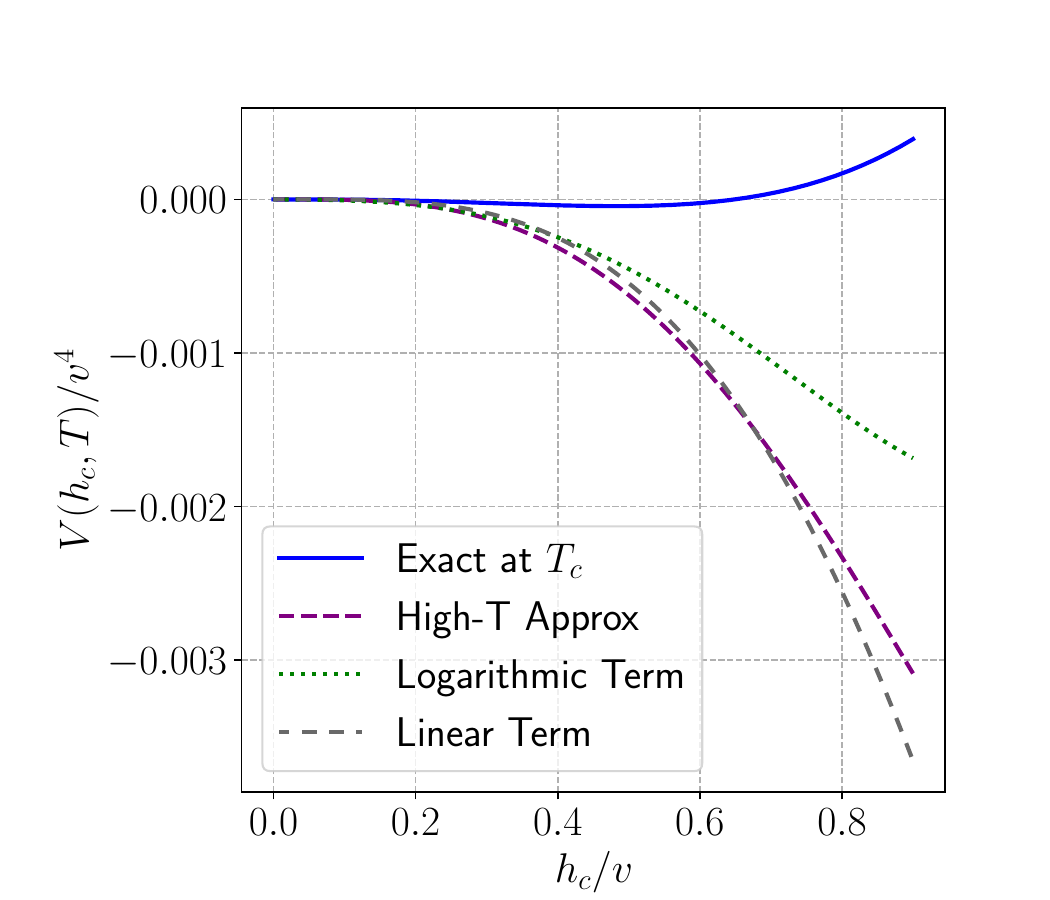}
    \includegraphics[width=0.43\linewidth]{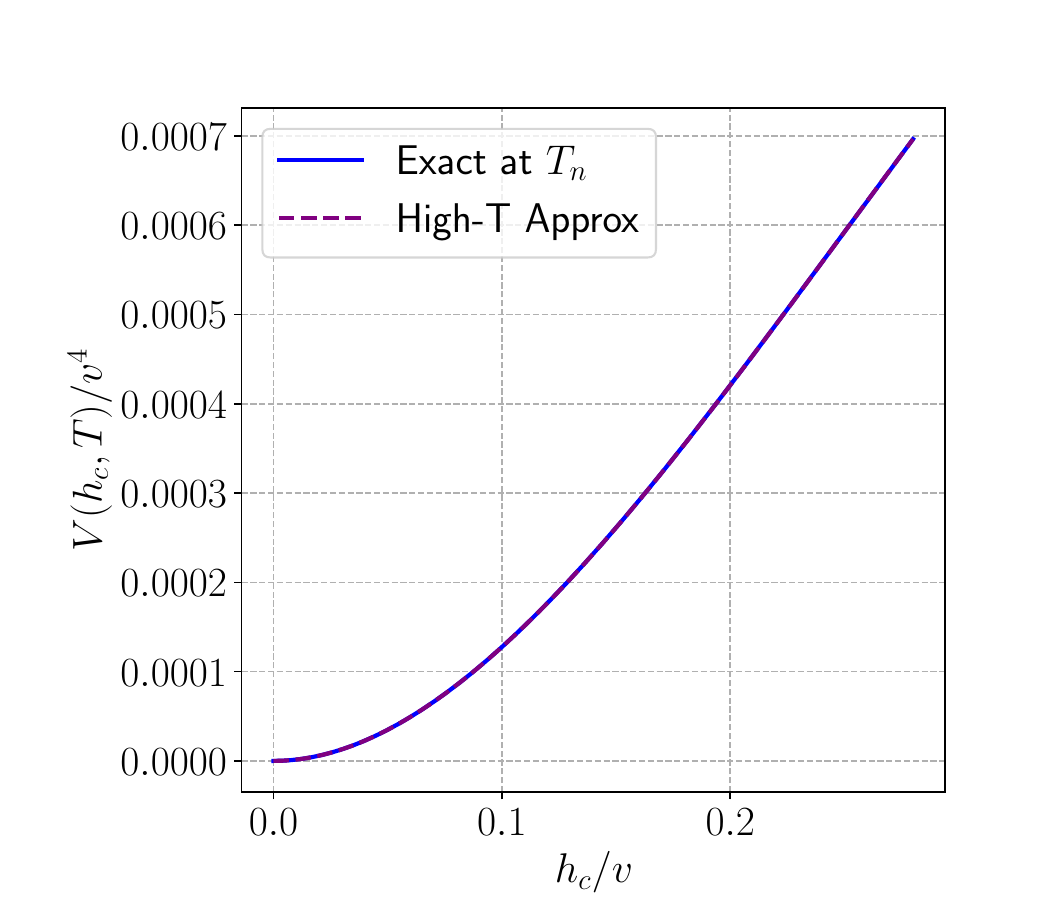}
     \includegraphics[width=0.43\linewidth]{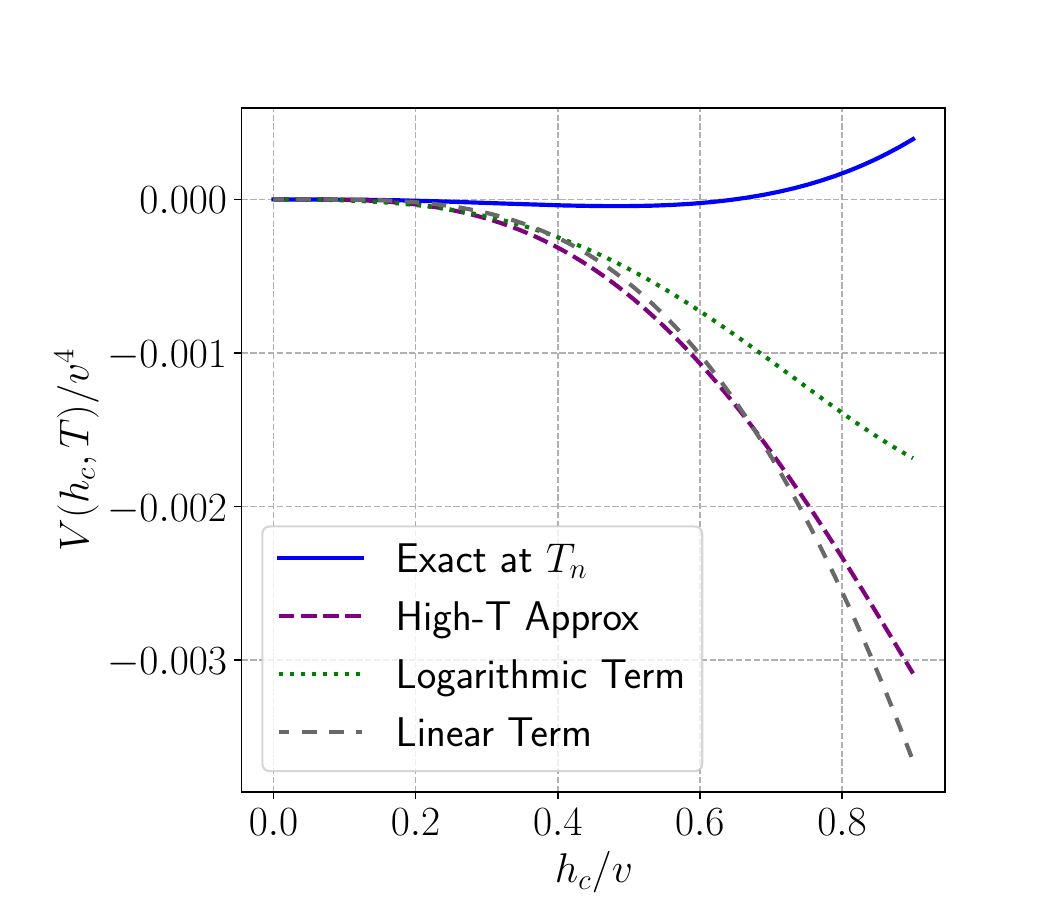}
    \includegraphics[width=0.43\linewidth]{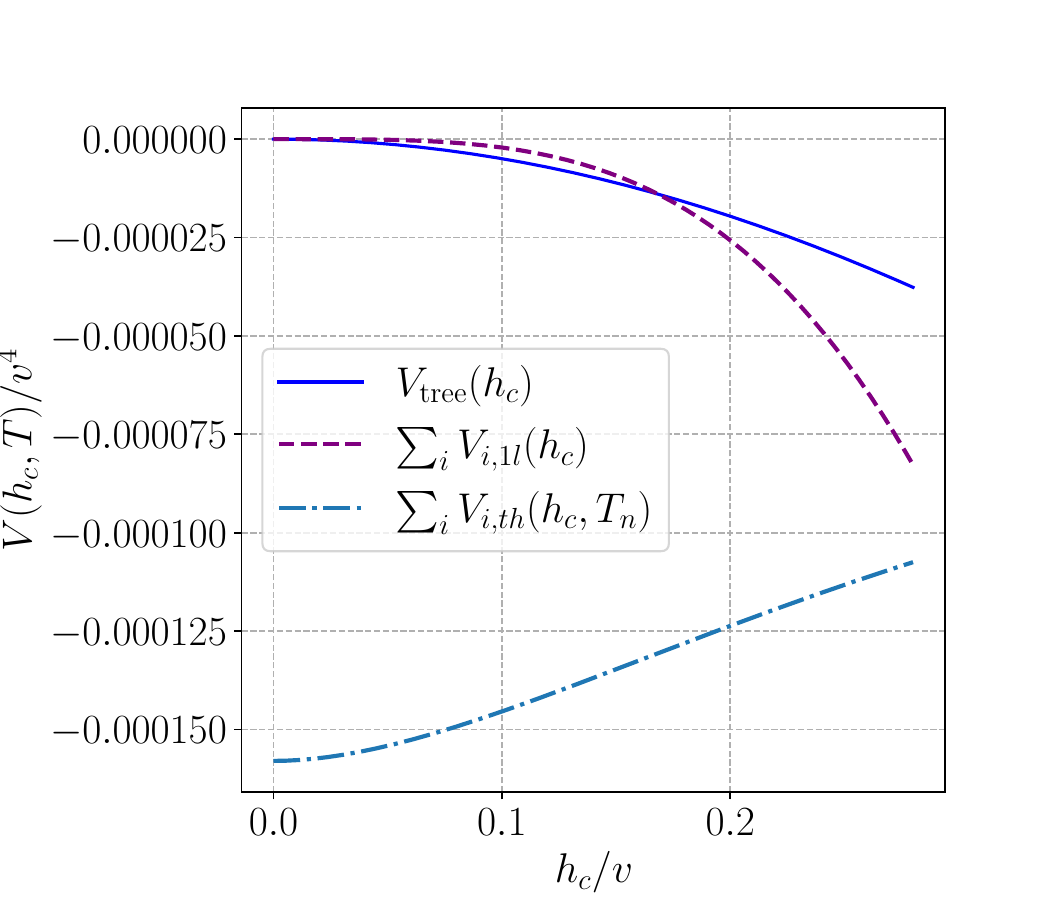}
     \includegraphics[width=0.43\linewidth]{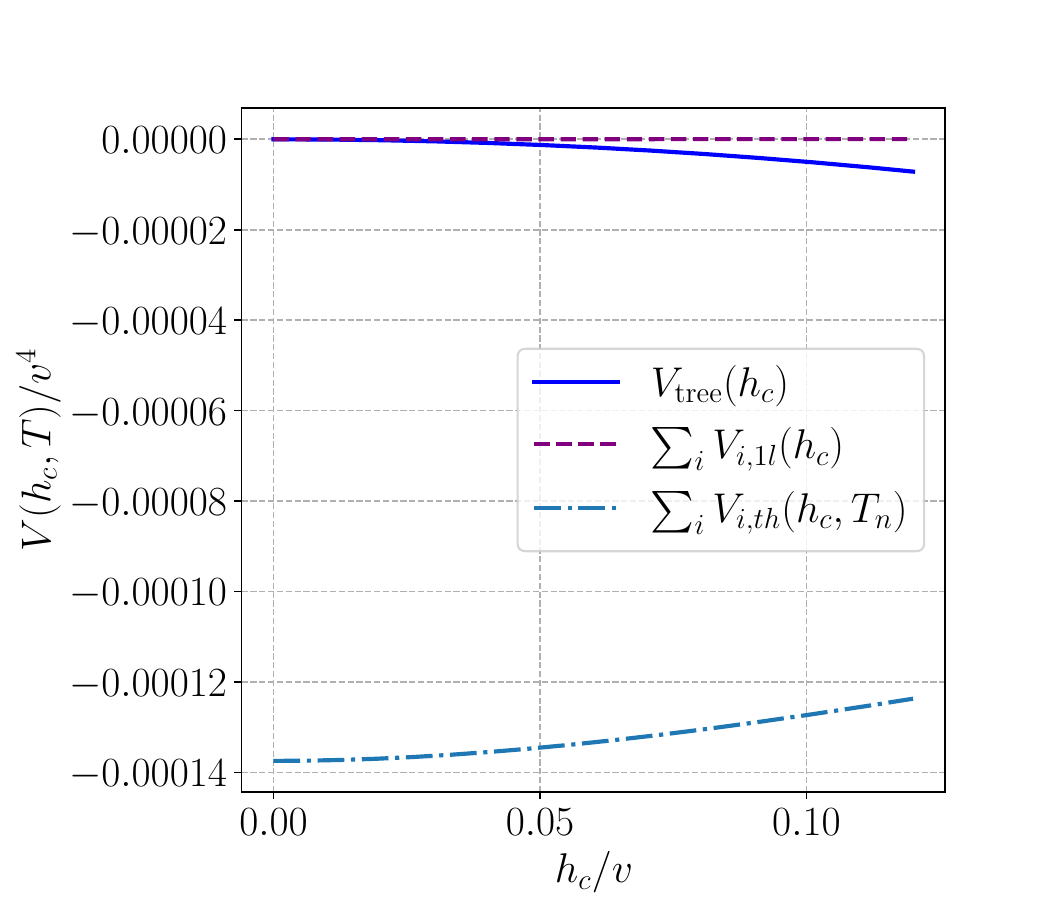}
    \caption{The effective potential $V(h_c,T)$, \eq{eq:completev1lu1}, including 1-loop corrections and thermal corrections. On the left-hand side we show the results for BP2 of Tab.~\ref{table:model_params}, and on the right-hand side we show the results of BP5, Tab.~\ref{table:params_wf}. BP2 corresponds to a case without fermions, while for BP5 one fermion pair is added.
    The top plots correspond to a comparison of the exact thermal corrections as given by the integral functions defining $J_b$ and $J_f$ in \eq{eq:effectivecorrpot} at $T_c$. The middle plots show the same comparison at $T_n$. The bottom plots show the individual contributions to the effective potential at $T_n$.  We see that while for the case of no fermions present in the model, the high-temperature approximation gives an idea of the exact behavior, when fermions are present instead this approximation is not good, as the temperature corrections are dominated by the logarithmic term and to some extent the linear term. }
    \label{fig:Potcomparisons}
\end{figure}

The computation of $\alpha$ and $\beta/H_*$ is affected by numerical uncertainties, both from intrinsic numerical algorithms and from the expression in \eq{eq:NTnO1}. In fact, varying the right-hand side of \eq{eq:NTnO1} between
$0.1$ and $10$, we find a variation of $T_n/v$, $\alpha$, and $\beta/H_*$ at the percent level. These uncertainties, however, are subdominant with respect to the value of the bubble wall velocity $v_w$.\footnote{While this manuscript was being finalized, the code \texttt{WallGo} was released \cite{Ekstedt:2024fyq}, which facilitates an accurate determination of the wall velocity and promises to reduce the uncertainty significantly in follow-up works.}
As we explain in more detail in App.~\ref{app:GWfor}, we choose to present our results with three values of $v_w$ to chronicle this uncertainty.

\section{Formulas for the gravitational wave density \label{app:GWfor}}
The density of GW is controlled by the following parameters: $T_n$, $\alpha$, $\beta$, the sound speed, $c_s$, the bubble wall speed $v_w$, and the efficiency factors $\kappa_\nu$, and $\epsilon$. 
In terms of these parameters, we can use the well-known formulas
\cite{Hindmarsh:2013xza,Hindmarsh:2015qta} for sound wave and turbulence contributions \cite{Caprini:2009yp,Binetruy:2012ze} when the barrier is mainly of thermal origin, that is, it disappears in the limit of $T\rightarrow 0$ and the 1-loop gauge coupling contributions are not negligible. This is the so-called case of \emph{non-runaway bubbles}. 
In this case, the main contributions come from the sound waves and the magnetohydrodynamic (MHD) turbulence of the plasma.
The red-shifted contribution to the GW density from sound waves today is given by 
\begin{multline} \label{eq:Omegasw}
\Omega_{\rsw} h^2 (f) =
\\
2.65 \times 10^{-6} \, H_*\tau_{\rsw}
\left(\frac{\beta}{H_*}\right)^{-1} v_w
\left(\frac{\kappa_\nu \alpha }{1+\alpha }\right)^2
\left(\frac{g_*}{100}\right)^{-\frac{1}{3}}
\left(\frac{f}{f_{\rsw}}\right)^3
\left(\frac{7}{4+3 \left(f/f_{\rsw}\right)^2}\right)^{7/2} ,
\end{multline}
where the peak frequency as observed today is given by
\bea
f_{\rm{sw}}=1.9\times 10^{-5} \, \frac{1}{v_w}
\left(\frac{\beta}{H_*}\right)\left(\frac{T_*  }{100 \, {\rm{GeV}}}\right)\left(\frac{g_*}{100}\right)^{1/6}{\rm{Hz}}, 
\eea
the factor $\tau_{\rsw}=\min\left[\frac{1}{H_*},\frac{R_*}{\bar{U}_f}\right]$ is
the time scale of the duration of the phase transition~\cite{Ellis:2018mja,Ellis:2019oqb},
and $g_*$ and $H_*$, respectively, are the number of degrees of freedom and the Hubble parameter at the time of GW production in the thermal bath. In the non-runaway bubble case, the reheating temperature and the thermal bath temperature, $T_*$, coincide with the nucleation temperature $T_n$.\footnote{Recalling that $T_{\rm rh}\approx T_n(1+\alpha)^{1/4}$ \cite{Ellis:2018mja}, so this is valid only for small values of $\alpha$.}
Note that $\tau_{\rm sw}$ can be either $1/H_*$ or ${R_*}/{\bar{U}_f}$, where
$H_*R_* = \max(v_w,c_s) \, (8\pi)^{1/3}(\beta/H_*)^{-1}$.
The root-mean-square (RMS) fluid velocity can be approximated as
\[
\bar{U}_f^2 \approx \frac{3}{4} \left(\frac{\kappa_\nu\alpha}{1+\alpha}\right).
\]
We recall that the simulations on which \eq{eq:Omegasw} is based were restricted to values of $\alpha \lesssim 0.1$ and $\bar{U}_f \lesssim 0.05$ \cite{Caprini:2015zlo, Caprini:2019egz}.
The efficiency factor $\kappa_\nu$ can be approximated by \cite{Espinosa:2010hh}
\bea
\kappa_\nu \simeq \left\{ 
\begin{array}{cc}
\alpha(0.73+0.83\sqrt{\alpha} +\alpha)^{-1}, & v_w \sim 1\,,\\
v_w^{6/5} 6.9 \alpha \left( 1.36-0.037 \sqrt{\alpha} + \alpha \right)^{-1}, &  v_w \ll 1 \,.
\end{array}
\right.
\eea

The MHD turbulence in the plasma is the sub-leading source of GW signals,
with the energy density being given by
\begin{multline} \label{eq:Omegaturb}
\Omega_\text{turb} h^2 (f) =
\\
3.35 \times 10^{-4}\left(\frac{\beta}{H_*}\right)^{-1}
v_w
\left(\frac{\epsilon\, \kappa_\nu \alpha }{1+\alpha }\right)^{\frac{3}{2}}
\left(\frac{g_*}{100}\right)^{-\frac{1}{3}}
\frac{\left(f/f_{\turb}\right)^3\left(1+f/f_{\turb}\right)^{-\frac{11}{3}}}{1+8\pi\frac{f}{h_*}}
\end{multline}
with
\begin{equation}
h_*=16.5 \frac{T_*}{10^8\, \rm{GeV}}\left(\frac{g_*}{100} \right)^{1/6} \text{Hz}
\end{equation}
and the peak frequency
\bea
f_{\rm{turb}}=2.7\times 10^{-5} \, \frac{1}{v_w}\left(\frac{\beta}{H_*}\right)\left(\frac{T_*}{100\,{\rm{GeV}}}\right)\left(\frac{g_*}{100}\right)^{1/6} {\rm{Hz}}.
\eea
We have assumed that the turbulence factor $\kappa_{\rm t}$ can be written as $\kappa_{\rm{t}}=\epsilon\kappa_\nu$, $\epsilon$ being another efficiency factor. The simulations in \cite{Hindmarsh:2015qta} suggest that only at most $5\%$ to $10\%$ of the bulk motion from the bubble wall is converted into vorticity, which enters into the turbulence contribution. Then it is customary to assume a conservative value of $\epsilon=0.05$ \cite{Caprini:2015zlo}.

The bubble wall velocity $v_w$ must be derived from a microphysical description of the interactions between the background scalar field evolving through the bubble wall and the thermal plasma. Since this process is in general quite challenging, we present the GW density profiles for three values of $v_w$: $0.1, 1.0$ and $v_w^d$. Here $v_w^d$ is the detonation velocity $v_w^d=(1/\sqrt{3}+\sqrt{\alpha^2+ 2 \alpha/3})/(1+\alpha)$ \cite{PhysRevD.25.2074}. The values for $\epsilon=0.05$, $\overline{U}_f\sim 0.05$ were obtained in the regime of detonations and small $\alpha$-deflagrations \cite{Hindmarsh:2015qta,Hindmarsh:2017gnf,Caprini:2015zlo,Caprini:2019egz}.

A further, subdominant, source of GW is the collision of bubbles.
An analytical expression based on simulations \cite{Cutting:2018tjt} is
\begin{multline}
\Omega_{\rm{c}} h^2 (f) =
\\
5.01 \times 10^{-6} \left(\frac{g_*}{100}\right)^{-\frac{1}{3}} v^2_w \, \left(\frac{\beta}{H_*}\right)^{-2} \left(\frac{ \kappa_{\coll} \alpha }{1+\alpha }\right)^{2}
\left(\frac{f}{f_{\rm col}} \right)^3
\left(1+2 \left(\frac{f}{f_{\rm col}}\right)^{2.07}\right)^{-2.18} .
\end{multline}
The efficiency factor $\kappa_\text{col}$ is more involved than other efficiency factors as it depends on the type of barrier and the model that produces the FOPT \cite{Cutting:2018tjt}.

Yet another subdominant contribution in the case of the non-runaway bubbles is the contribution from the scalar field.  The expressions of this contribution have been estimated using simulations in the envelope approximation \cite{Kamionkowski:1993fg}  and in the thin-wall and envelope approximations \cite{Huber:2007vva}. Another way is to compute it analytically in terms of the two-point correlator of the energy-momentum tensor \cite{Jinno:2016vai}, 
which yields
\bea
\Omega_{\phi} h^2 (f_{\text{peak}}) \simeq
1.67\times 10^{-5} \, \kappa_\phi^2 \Delta_{\text{peak}}
\parent{\frac{\beta}{H_*}}^{-2}\parent{\frac{\alpha}{1+\alpha}}^2\parent{\frac{g_*}{100}}^{-\frac{1}{3}} ,
\eea
where
\be
\Delta_{\text{peak}}=\frac{0.48v_w^3}{1+5.3v_w^2+5.0v_w^4} \,.
\ee

Both the bubble collision and the scalar field contributions are proportional to $(\beta/H_*)^{-2}$ and therefore suppressed with respect to the sound wave and turbuluence contributions.
Hence, we have not included these contributions in our numerical analysis.

\bibliographystyle{JHEP}
\bibliography{ref}

\end{document}